\documentclass[letterpaper, 10pt, twocolumn, twoside]{article}
\usepackage[hmargin=0.7in,vmargin=1in]{geometry}

\usepackage[utf8]{inputenc}
\usepackage[T1]{fontenc}
\usepackage{titling}
\usepackage[affil-it]{authblk}

\usepackage[style=nature, backend=biber,sorting=none]{biblatex}

\AtEveryBibitem{\clearfield{url}}
\DeclareFieldFormat{url}{}

\usepackage{float}
\usepackage{graphicx}
\usepackage{color,soul}
\usepackage[dvipsnames]{xcolor}

\usepackage[dvipsnames]{xcolor}
\usepackage{setspace}

\usepackage{mathtools}
\usepackage{amsmath}
\usepackage{bm} 

\usepackage{blindtext}

\title{A Kerr soliton Ising machine for combinatorial optimization problems}

\author[1,2]{Yan Jin \thanks{yan.jin@nist.gov}}
\author[1,2]{Nitesh Chauhan}
\author[1,2]{Jizhao Zang}
\author[3]{Brian Edwards}
\author[3]{Pratik Chaudhari}
\author[3]{Firooz Aflatouni}
\author[1,2]{Scott B. Papp}
\affil[1]{Time and Frequency Division, National Institute of Standards and Technology, Boulder, CO USA}
\affil[2]{Department of Physics, University of Colorado, Boulder, CO, USA}
\affil[3]{Department of Electrical and Systems Engineering, University of Pennsylvania, Philadelphia, PA, USA}

\date{}

\pretitle{\begin{center}\large\bfseries}
\posttitle{\par\end{center}}
\preauthor{\begin{center} \large}
\postauthor{\end{center}}
\predate{\begin{center} \normalfont}
\postdate{\end{center} \vspace{-10pt}}

\begin{document}
\maketitle

\textbf{\noindent The growing challenges of scaling digital computing motivate new approaches, especially through the dynamical evolution of physical systems that mimic neural networks and combinatorial optimization problems  \cite{mohseni2022ising,tatsumura2021scaling,takesue2025finding}. While light is a hyper efficient information carrier, intrinsically weak light interactions make direct information processing difficult to implement \cite{mcmahon2023physics}. Recently, specialized nonlinear photonics have opened new controls over light fields with extraordinary bandwidth, coherence, and the emergence of strong interactions among nonlinear eigenstates like solitons \cite{cole2017soliton}. We harness an ensemble of hundreds of Kerr-nonlinear microresonator solitons and implement an analog feedback network to create an Ising machine with fully programmable all-to-all interactions. By increasing the feedback for self, on-diagonal interactions, each soliton exhibits a universal spin-like bifurcation. Using this palette of interactions amongst the entire soliton ensemble, we encode the Ising machine to solve the benchmark Boolean satisfiability problem (SAT). The combination of uniform soliton interactions and the compatibility of our Ising machine with high-speed data interconnects enables rapid and precise solutions of complex SAT problems. Indeed, the soliton properties bound the tradeoff of optical power and time use by the machine at approximately 10 mW and 1 $\mu$s for a single feedback step. We performed >10,000 trials on more than 100 randomly generated SAT instances to evaluate the Ising machine, demonstrating the potential to exceed the performance of benchmark digital SAT solvers. Our work highlights the convergence of optical nonlinearity, ultralow loss photonics, and optoelectronic circuits, which can be combined for a wide range of computation-acceleration tasks.
}

The end of rapid scaling in digital-computer technology and the rise of machine optimization and artificial intelligence has opened interest in controllable and scalable physical systems to solve combinatorial optimization problems, which are widely used in schedule optimization, finance, chemical discovery, and deep neural network architectures \cite{mohseni2022ising, mizuno2024finding, laydevant2024training, sharma2023augmenting}. In particular, physical approaches leverage quantum-inspired behaviors like discrete spin interactions, featuring sharp state transitions and the potential for scalable spin networks with low operation power. Since combinatorial optimization can fall into the NP-complete category, they are challenging to solve with digital computers. Boolean satisfiability (SAT) is the canonical NP-complete example, meaning that every problem in NP can be efficiently reduced to SAT, and thus SAT serves as a central benchmark. 
Therefore, numerous non-von-Neumann architectures have been explored to solve these problems by mapping them to an Ising model. An extraordinary range of physical systems can be arranged to create an Ising machine, notably parametric oscillators \cite{doi:10.1126/science.aah4243, doi:10.1126/sciadv.abh0952, yamamoto2017coherent}, opto-electronic networks \cite{bohm2019poor, bohm2022noise}, Kerr nonlinear optics \cite{quinn2024coherentisingmachinebased}, electronic circuits \cite{sharma2023augmenting, pedretti2025solving}, and CMOS circuits with static random-access memory (SRAM) \cite{dee2025108mwmixedsignalsimulatedbifurcation}.

Despite their connection with interacting quantum-spin systems, Ising machines are implemented with a range of architectures in which memory, connectivity, interaction speed, and scalability are key. A notional digital/analog electronic circuit can implement an Ising machine in which pristine spin states are stored in a digital SRAM array. Connectivity is realized through multiplexing of analog current signals derived from the SRAM. Nonlinearity is introduced via comparators that project the analog interaction sum back into discrete digital spin states, closing the feedback loop.  However, the power consumption of digital-to-analog and analog-to-digital converters, latency in SRAM access, and signal propagation delay in comparators impose fundamental limits on update throughput and ultimately constrain the overall spin capacity and performance of such hybrid systems. These ubiquitous CMOS constraints motivate the exploration of alternative physical substrates like photonics that promise lower energy per operation and far greater parallelism.

\begin{figure*}[ht!]
	\centering%
	\includegraphics[width=0.85\textwidth]{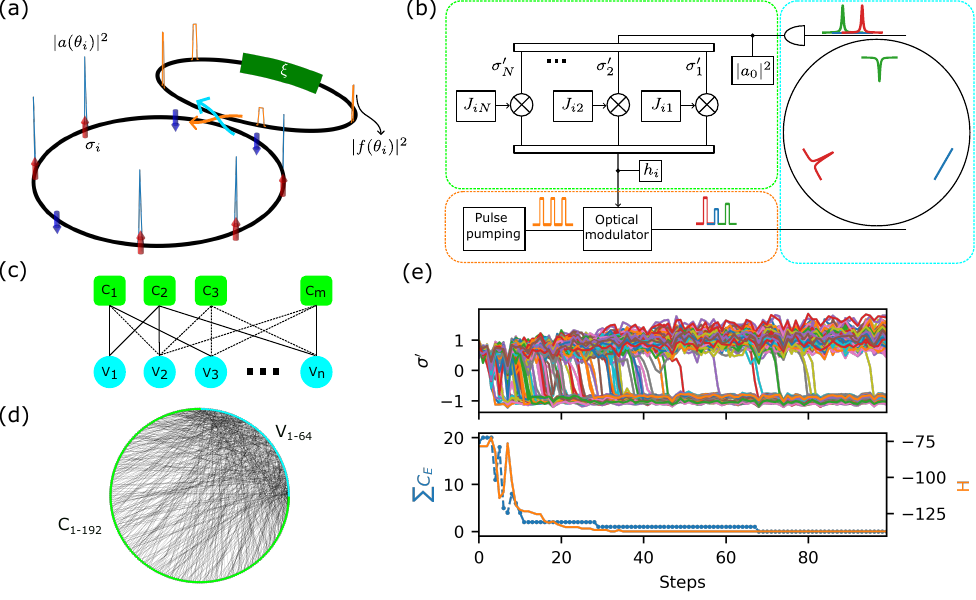}%
	\caption{Kerr soliton Ising machine. (a) The resonator with soliton spins (circle) and the feedback network with the pump and the feedback circuit (oval). Pump pulses (orange) and solitons (blue) are described by $|f(\theta_i)|^2$ and $|a(\theta_i)|^2$ in the spatial domain, respectively, and $\xi$ refers to the feedback circuit.
     (b) Schematic of the Ising machine, including the feedback circuit (green box), the soliton resonator and photodetection (blue box), and the pump (red box). 
     (c) Structure of a 3-SAT problem, where $V_i$ ($i\in \{1,2,... n\}$) and $C_i$ ($i\in \{1,2,... m\}$) are variables and clauses, respectively. The solid lines connect variables, and dashed lines connect the negation of variables. 
     (d) A 3-SAT problem with 64 variables and 192 clauses. (e) Evolution of the Ising machine versus gradient-descent step for the 3-SAT problem in (d). In the upper panel, the soliton spins bifurcate to 2 states. In the lower panel, the number of unsatisfied clauses $\sum C_E$ and the Hamiltonian $H$ evolve to the global minimum. }\label{fig1}
\end{figure*}

Kerr-microresonator solitons are discrete and localized electromagnetic excitations of an intraresonator field, exhibiting dissipation, gain, and the balance of nonlinearity and phase matching \cite{herr2014temporal, tobiasreview2018, leo2010temporal}. Such solitons exhibit modelocking and they form an outcoupled, ultrafast optical pulse train from the resonator that we associate with an optical frequency comb. Hence, we refer to this system as a soliton microcomb, and it exhibits extraordinary optical and microwave coherence with fractional optical frequency accuracy and precision reaching $10^{-20}$ \cite{drake2019terahertz}, which supports applications ranging from optical frequency metrology \cite{drake2019terahertz} to optical data interconnects to universal electronic synthesis \cite{zang2025universalelectronicsynthesismicroresonatorsoliton}. From the earliest experiments with Kerr solitons, their discreteness was recognized, and ensembles of solitons feature a high degree of indistinguishability and long-range interactions \cite{cole2017soliton}. Moreover, soliton microcombs can be implemented with increasingly advanced photonic integration technologies \cite{spott2017heterogeneous} and foundry manufacturing \cite{liu2025implementing}.

Here, we introduce a Kerr-soliton Ising machine and use it to solve combinatorial optimization problems. We implement the Ising machine with a large ensemble of Kerr solitons in a resonator, and we map fully programmable all-to-all interactions amongst the ensemble with an opto-electronic feedback network. The emergent collective state of the discrete solitons represents the spins of the Ising model, and we program the hybrid opto-electronic feedback network to directly couple SAT instances, multiplexing the solitons in time and amongst one electrical path for each optimization problem mapped to the Ising model. 
The soliton spin bifurcation arises directly from Kerr nonlinearity because there are only two, distinct, quantum-based eigenstates of the resonator \cite{chembo2016quantum}, the empty state and the soliton state.
To evolve the Ising machine, we photodetect the soliton spin states and use stepped gradient descent to update the feedback circuit. Each step fundamentally requires 3.5 $\mu$s of time in accord with the soliton nonlinear dynamics. Moreover, each soliton only consumes around 2.3 mW of pump power in the present apparatus. Hence, we can solve complex, arbitrary SAT problems more rapidly and with two orders of magnitude less optical energy than benchmark digital computer algorithms. Furthermore, the Kerr soliton Ising machine shows low latency, rapid execution, energy efficiency, and good stability and scalability.

The concept of our Kerr soliton Ising machine is shown in Fig. \ref{fig1}(a), including a high finesse Kerr resonator (circle) to generate solitons and a low finesse feedback network (oval) to mediate all-to-all soliton interactions. Kerr solitons propagate endlessly in a resonator without temporal spreading, and each spatial coordinate of the resonator can support at most one soliton. Moreover, as we increase the pump power there is a sharp transition from the empty state to the soliton.
Each of the solitons (blue) is excited by the corresponding pump pulses (orange) coupled from an external path. We model the behavior of solitons with the Lugiato-Lefever equation (LLE) \cite{godey2014stability}, including use of the Kerr soliton Ising machine to solve 3-SAT problems; see Methods. The intracavity field $a(\theta_i)$ for each of the $N$ solitons is controlled by a repeating sequence of incoming pump pulses $f(\theta_i)$, where $\theta_i$ ranges from $2\pi i/N$ to $2\pi(i+1)/N$. 
The LLE solutions indicate that the two states of the solitons, i.e. either excited or empty are two nonlinear eigenstates, and the blue traces on top of a circular ring in Fig. \ref{fig1}(a) show the multi-soliton ensemble in a resonator mapped into mathematical spins in the Ising model. The Hamiltonian or Ising energy is
\begin{equation}
    H=\bm{\sigma}^\mathsf{T}\cdot \bm{J}\cdot\bm{\sigma} + \bm{h}\cdot\bm{\sigma} = \sum_{i,j}J_{ij}\sigma_i\sigma_j + \sum_i h_i \sigma_i, \label{eq1}
\end{equation}
where $\bm{\sigma}$ is the Ising spin array with $\sigma_i\in \{-1, 1\}$ being the $i$-th spin, $\bm{J}$ is the interaction matrix with $J_{ij}$ being its element, and $\bm{h}$ is the external force array with $h_i$ the elements. Inside the resonator, the spin is up ($\sigma_i=1$, marked with red arrows) if the soliton is excited and the peak of its power within $[2\pi i/N, 2\pi(i+1)/N]$ reaches a given value $|a_0|^2$, and down ($\sigma_i=-1$, marked with blue arrows) otherwise.

While the solitons provide a highly robust physical platform to mimic spin behavior, the operational key to information processing in the Ising machine is the external feedback network. Outcoupled solitons are multiplexed and modulated with a time-domain signal through a feedback circuit ($\xi$ in Fig. \ref{fig1}(a)) to program $\bm{J}$. The feedback circuit acts through the intensity of the pump pulses, and we present its architecture in more detail in Fig. \ref{fig1}(b) with the feedback circuit and the pump laser shown in green and orange boxes, respectively. 
In the feedback circuit, we realize hybrid multiplexing of solitons through propagation in $N$ paths and by application of time dependent attenuation of the paths. Operationally, we photodetect the soliton intensity $|a(\theta_i)|^2$ from the drop port of the resonator, apply a preset bias $|a_0|^2$ to bifurcate two soliton spin levels $\sigma_i'\propto |a_i|^2-|a_0|^2$, where $|a_i|^2$ is the peak of $|a(\theta_i)|^2$; see Methods. While mathematical spins $\sigma_i$ are discrete, $\sigma'_i$ are continuous and range approximately from -1 to 1.
To drive the Ising machine system towards its minimum energy, we perform a gradient-descent update on each spin. The feedback term has the form of $\xi_i = \eta(-2\sum_j J_{ij} \sigma_j' - h_i)$ from Eq. \ref{eq1}, where $\eta$ denotes an overall attenuation factor.
To form $\xi_i$ for the soliton spin at site $i$, we multiply each of $\sigma_j'$ with the fixed interaction term $2J_{ij}$ in each path, combine them and bias the combined signal with $h_i$. We generate an electric waveform with $\xi_i$ and apply it to the optical modulator in the same order as the soliton sequence in the time domain. With time and path multiplexing, the feedback network realizes all-to-all interactions and individually addresses the solitons at each site. The size of the feedback network scales linearly with the number of soliton spins $N$, indicating that our system can scale up to address even larger optimization problems.

We coordinate operation of the solitons and the feedback network with the pump laser and the photodetector; see the blue box in Fig. \ref{fig1}(b). We generate temporally broad pump pulses (orange), and then adjust the intensity of each pulse to be $|f(\theta_i)|^2=|f_0(\theta_i)|^2(1+\xi_i)$ with the optical modulator, where $f_0(\theta_i)$ is chosen such that $|a_i|^2 \simeq |a_0|^2$; see Methods. All the pump pulses couple to the Kerr resonator, and we characterize $|f(\theta_i)|^2$ in terms of the threshold power to generate a soliton. The solitons in the Kerr resonator evolve at the timescale of the photon lifetime, which is 0.3 $\mu$s,
and the soliton ensemble is efficiently outcoupled from the resonator for photodetection.

We benchmark the performance of the Kerr-soliton Ising machine by solving arbitrary, randomly generated SAT problems. Boolean satisfiability is a decision problem about whether a Boolean formula can be satisfied by assigning truth values to its variables.
A problem (Fig. \ref{fig1}(c)) consists of $n$ variables $V_i$ ($V_i\in\{\mathrm{True}, \mathrm{False}\}$) and $m$ clauses $C_i$ ($C_i\in\{\mathrm{True}, \mathrm{False}\}$), which are the logical OR of the variables or their negation.
Specifically, we solve 3-SAT problems in which each clause contains three variables or their negation. Figure \ref{fig1}(c) presents an example 3-SAT problem with the variable nodes along a blue circle and the clause nodes in a green square. We use the solid line when a variable is in a clause, and the dashed line when the negation of a variable is in a clause. 
For example, the clause node $C_1$ is connected to $V_1$ and $V_3$ with solid lines, and to $V_2$ with dashed lines, thus can be written as $C_1 = (V_1 \lor \neg V_2 \lor V_3)$, where $\neg V_2$ is the negation of $V_2$. If $V_1$ and $V_3$ are false and $V_2$ is true, then clause $C_1$ is unsatisfied. We define the number of unsatisfied clauses as $\sum C_E$. To solve the 3-SAT problem with the Ising machine, we first translate it into a quadratic unconstrained binary optimization (QUBO) problem \cite{QUBO} and then convert it to an Ising model with an efficient algorithm on a digital computer. A 3-SAT problem with $n$ variables and $m$ clauses can be transformed into an Ising problem with $N=m+n$ spins, and finding the ground state of the Ising problem is equivalent to finding the solution to the 3-SAT problem, i.e., a set of $V_i$ such that the Boolean formula $C_1\wedge C_2 \wedge ... C_m$ is satisfied and $\sum C_E=0$.

Figure \ref{fig1}(d) demonstrates use of the Kerr soliton Ising machine to solve a 3-SAT problem that contains $n=64$ variables and $m=192$ clauses. We translate this problem to an Ising model with $N=256$ spins, and encode the interaction matrix $\bm{J}$ and external field $\bm{h}$ to the feedback circuit. 
The computation begins with a setting $\eta=0$ and injection of a train of pump pulses with uniform intensity. We photodetect the soliton spins $\sigma_i'$ to obtain the feedback term $\xi_i$ and modulate the peak power of pump pulses with $\xi_i$, which completes the first step of gradient descent. 
Then we gradually increase $\eta$ until the Ising machine runs a certain number of steps $N_s$ to reach the final state. 
The choice of $N_s$ is crucial, since sufficient steps are needed to reach a stable minimum of the Ising energy. 
For this problem, we take $N_s=100$ steps to drive the Ising machine to the final state.

\begin{figure}[ht!]
	\centering%
	\includegraphics[width=\columnwidth]{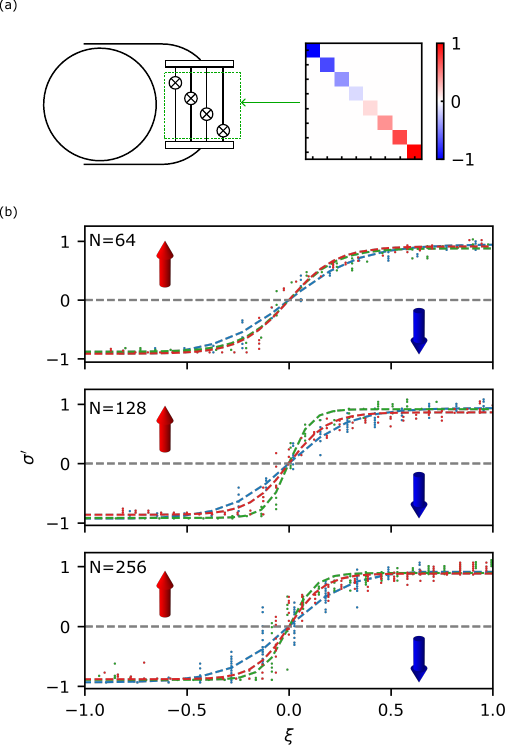}%
	\caption{(a) Feedback with self interaction. The diagonal $\bm{J}$ matrix is sent to the feedback circuit of the Ising machine system to realize the self interaction. (b) Nonlinearity with $N$= 64, 128 and 256 soliton states in one round trip. For each $N$ we set different feedback ranges $\xi\in [-1, 1]$ (green), $[-0.8, 0.8]$  (red) and $[-0.7, 0.7]$ (blue). The dots are measured data points and the dashed lines are the fitting. }\label{fig2}
\end{figure}

Figure \ref{fig1}(e) presents the evolution of the spin values $\sigma'$, the corresponding Ising energy $H$, and the number of unsatisfied clauses $\sum C_E$ with each step. 
As we increase $\eta$, the soliton spins are strongly projected to two, distinct stable states -- either excited or empty, and $H$ and $\sum C_E$ converge to a minimum. As $\sum C_E$ reaches zero, all the clauses are satisfied for a complete solution of the 3-SAT problem. Gradient descent applied to the Ising model is inherently probabilistic, as the non-convex energy landscape of the Ising interaction leads to local minima and no guarantee of convergence to the global minimum. Therefore, we perform repeated trials of the machine, testing each trial for $\sum C_E=0$. The soliton amplitude bifurcation that enables our Ising machine arises from feedback-induced bistability of the soliton field due to Kerr nonlinearity, providing a physical representation for discrete spin states. The machine works for a wide range of $\eta$, as sharp soliton state transitions (Fig. \ref{fig1}(e)) occur with the increase of $\eta$ from zero. 
This result aligns with the prediction from the LLE \cite{godey2014stability}; see Methods.

To investigate the soliton spin bifurcation, we program an on-diagonal self-interaction Hamiltonian ($J_{ij} = J_{ii} \delta_{i,j}$) with off-diagonal terms zero, as shown schematically in Fig. \ref{fig2}(a). 
We vary the diagonal terms $J_{ii}$ from -1 to 1, which effectively creates a train of pump pulses with increasing intensity.
To explore the response under different conditions, We select $N$=64, 128 and 256 soliton states in a round trip, as shown in Fig. \ref{fig2}(b), and for each $N$, we vary $\eta$ to prepare three different self-interacting Hamiltonians to generate the feedback $\xi_i$ within 3 different ranges: $\xi_i \in [-1, 1]$, $\xi_i \in [-0.8, 0.8]$ and $\xi_i \in [-0.7, 0.7]$.
These ranges correspond to the green, red, and blue datasets in Fig. \ref{fig2}(b). 
For each experimental setting of $\xi_i$, we measure $\sigma'$ (dots) and fit the data with the sigmoidal function (dashed line). We adjust $|a_0|^2$ to ensure that the soliton state  reaches the threshold when the feedback $\xi=0$ (gray dashed line), and assign spin-up ($\sigma_i=1$) in the red arrow region or spin-down ($\sigma_i=-1$) in the blue arrow region based on the sign of $\sigma_i'$. 
The soliton states $\sigma'$ present a slightly stronger bifurcation with larger $\eta$
, which likely results from saturation of the photodetector and EDFA. In our experiments, the observed bifurcation predominantly originates from the Kerr nonlinearity of the soliton field. The universal bifurcation behavior of the solitons across a wide range of pump powers and experimental conditions ensures robust spin-state discrimination, serving as a key enabler for efficient and scalable SAT problem solving in the Kerr-soliton Ising machine.

\begin{figure*}[t!]
	\centering%
	\includegraphics[width=0.85\textwidth]{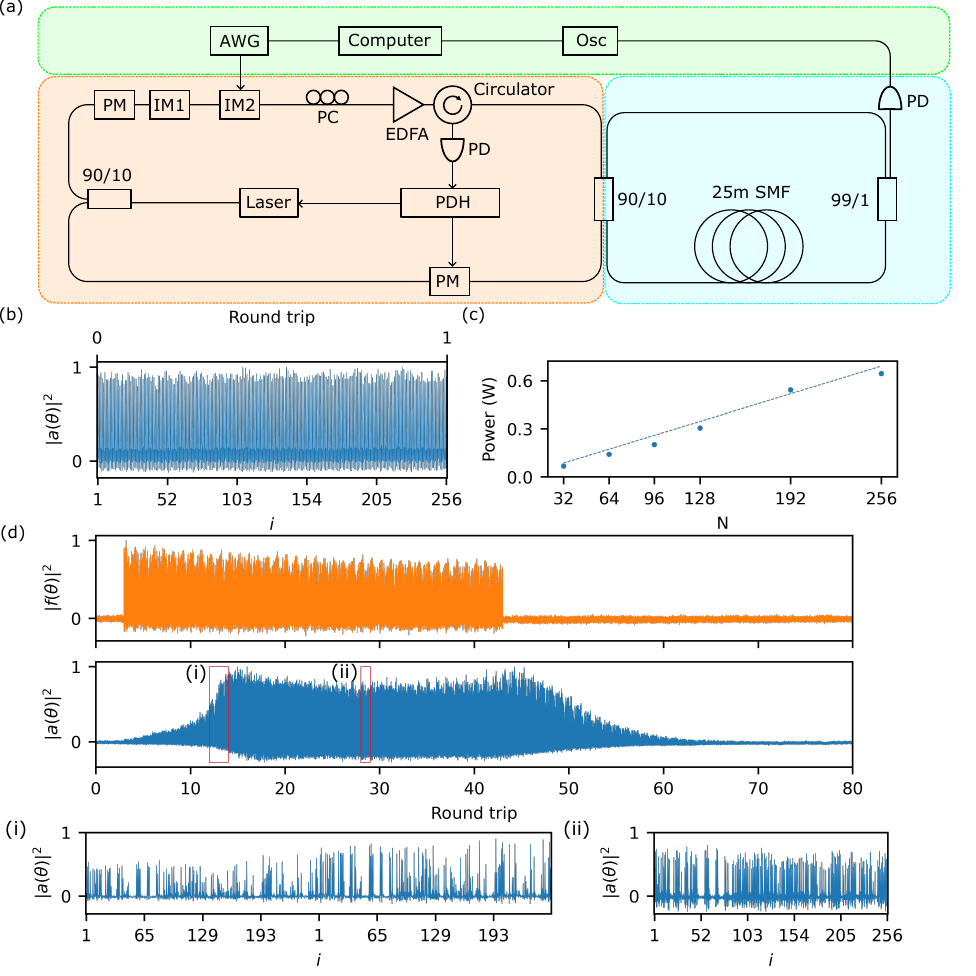}%
	\caption{(a) Setup. PM: phase modulator. IM: intensity modulator.
        AWG: arbitrary waveform generator.
        Osc: oscilloscope.
        PD: photodetector .
        PDH:  Pound–Drever–Hall locking.
        EDFA: Erbium-doped fiber amplifier.
        PC: polarization controller.
        SMF: single-mode fiber.
        (b) The intracavity field intensity $|a(\theta)|^2$ for $N=256$ solitons in a round trip.
        (c) Minimal EDFA power required to generate solitons for different $N$.
        (d) Pump pulses (orange) and solitons (blue) in a step. The pump is on for 40 round trips and off for another 40 round trips. The corresponding solitons evolve to a stable state and fade after the pump is off. Stage (i) presents 2 round trips of solitons during initial build-up, and stage (ii) presents 1 round trip of solitons at the stable state.
        }\label{fig3}
\end{figure*}

We present the experimental setup of the soliton Ising machine in Fig. \ref{fig3}(a). The pump laser, feedback circuit, and the Kerr resonator are highlighted in orange, green, and blue, respectively.
For the pump, a 1556 nm laser is the source for Pound–Drever–Hall (PDH) locking and pulse-pump generation, using an electro-optic frequency comb driven at 7.4 GHz by a phase modulator (PM) and an intensity modulator (IM1) \cite{drake2019terahertz, Zang:25}.
We use a separate intensity modulator (IM2) to close the feedback circuit from the PD to the pump pulse intensity. Operationally, in the feedback circuit we photodetect the soliton intensities, digitize $|a_i|^2 - |a_0|^2$, and use a computer to generate an electrical waveform with 20 GHz bandwidth that applies $\xi_i$; see Methods. 
The pump pulses are amplified by an erbium-doped fiber amplifier (EDFA) and injected into the fiber resonator. 
We use a circulator to measure the reflected power for PDH locking. 
We generate solitons in a 25 meter single-mode fiber (SMF) resonator with a round trip of $T_r=138$ ns. 
The resonator incorporates two SMF couplers: one with coupling ratio of 90/10 for injecting the pump pulses into the resonator and one with 99/1 ratio for extracting the soliton spins and monitoring the PDH locking state.
We pump the resonator with a train of pulses with linearly increasing intensity as in Fig. \ref{fig2} to record the site of each soliton state and calibrate $|a_0|^2$ i with $\sigma'\simeq 0$ when $\xi=0$.
During each gradient-descent step of a machine run, we calculate the feedback $\xi_i=\eta(-2\sum_j J_{ij}\sigma_j'-h_i)\propto -\mathrm{d} H/ \mathrm{d} \sigma_i$ and update the new waveform of $f(\theta_i)$, which precisely mimics the behavior of a path-multiplexed electronic circuit.

Scalability of the soliton spin number is an important advantage of our Ising machine, since this enables solution of larger optimization problems. Figure \ref{fig3}(b) demonstrates generation of a $N=256$ soliton ensemble within one round trip of the resonator, leveraging 256 independently programmable pump pulses. 
The average power of the pulsed pump laser naturally scales with $N$, which we characterize in Fig. \ref{fig3}(c). We measure the minimum, threshold optical power required to create the soliton ensemble as we vary $N$, demonstrating that 2.3 mW, which is comparable to the predicted threshold 0.9 mW; see Methods. In our experiments when solving 3-SAT problems, we set the EDFA power to twice the threshold.


The rapid evolution of Kerr solitons is a key feature of the Ising machine.
We record the full dynamics of the solitons in the time domain with the oscilloscope when solving an Ising problem with $N=256$ spins, as shown in Fig. \ref{fig3}(d).
We repeatedly send 80 round trips of pump waveforms $|f(\theta)|^2$ (orange) at each gradient descent step, where the pump pulses in the first 40 round trips are replicated to continuously excite the solitons at the corresponding sites, and the waveform in the last 40 round trips is null to maintain a buffer between the rapidly repeated runs of the machine; see Methods.
We measure the soliton waveform $|a(\theta)|^2$ (blue) and align the start of the pump and soliton waveforms.
In the beginning, the soliton intensity is low. 
The solitons gradually evolve to a steady state with continuous pulse pumping, and fade away as we turn off the pump. 
We present 2 stages of solitons during a single step: 2 round trips of initial build-up in stage (i) and 1 round trip of the stable state in stage (ii).
As the system evolves, the peak power of the solitons in each round trip gradually increases until they reach a constant value.
It requires more time to accumulate the power in the fiber resonator and generate solitons for some pump pulses with lower power.
Therefore, the stable state of stage (ii) contains more solitons, despite the fact that the peak power in stage (i) is comparable to that in stage (ii).
The operation time during which the solitons become stable depends fundamentally on the external coupling rate, which we can control by changing the couplers with different coupling ratios.

\begin{figure*}[t!]
	\centering%
	\includegraphics{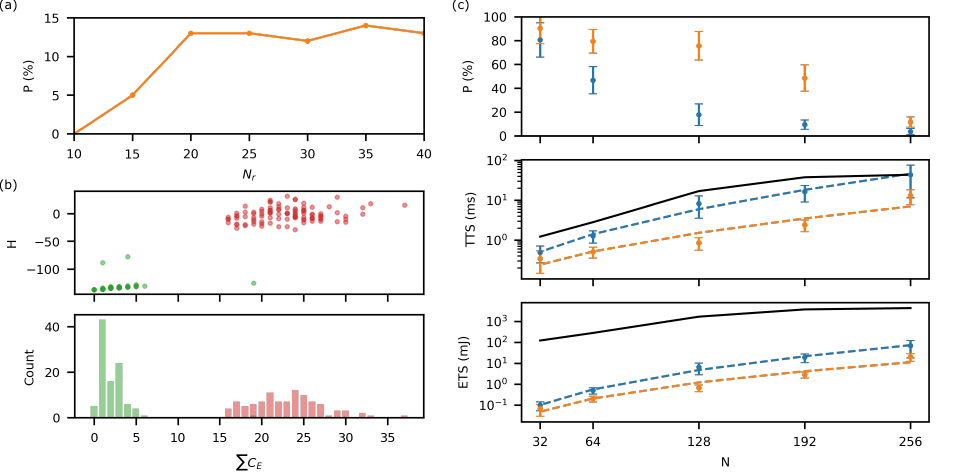}%
	\caption{(a) Success probability $P$ with different $N_r$. (b) Ising energy landscape and distribution of number of unsatisfied clauses $\sum C_E$. The red dots indicate the Ising energy $H$ and number of unsatisfied clauses $\sum C_E$ in the initial state, and the red bars indicate the initial conditions. The green dots and bars are for final states. (c) The success probability ($P$), $\rm{TTS}$ and $\rm{ETS}$ for different numbers of spins $N$ and for different problem difficulty (orange for $m/n=2$, and blue for $m/n=3$). The dashed lines are the fitting, and the error bars denoting one standard deviation of the data are calculated from results of 5 different problems for each $N$. The black curves are results using the WalkSAT algorithm.}\label{fig4}
\end{figure*}

We explore use of the Kerr soliton Ising machine to solve arbitrary, randomly generated 3-SAT problems, benchmarking the machine in terms of unit-normalized solution probability ($P$) and the time to solution ($\rm{TTS}$); see Fig. \ref{fig4}. Since 3-SAT is NP-complete, there do not exist polynomial-time deterministic solvers, and probabilistic approaches, e.g. our Ising machine and the widely used benchmark WalkSAT algorithm \cite{selman1993domain, selman1994noise}, require repeated trials to search the vast solution space. First, we characterize the dependence of $P$ on $N_r$, which represents the amount of time per gradient-descent step that the solitons are active in problem solution. For a 3-SAT problem with $n=85$ and $m=171$, we map it to the Ising Hamiltonian with $N=256$ with the procedure in \cite{QUBO} by use of a digital computer. Similar to the procedure of Fig. \ref{fig3}(d) for operation of the Ising machine, we define a repeated pump-pulse sequence of 80-$T_r$ duration with 50\% duty cycle that initiates solitons and evolves the system with the $J_{ij}$ interaction. Repeated trials of the pulse sequence search for solutions randomized by fluctuations in the solitons.  To characterize $P$, we vary the setting of $N_r$, which defines the evolution time of the soliton intensities $|a(\theta_i)|^2$ for each step of gradient descent. The solution probability clearly saturates, and we fix $N_r = 25$ for all of the experiments reported in this paper. This corresponds to a time of $T_s=N_rT_r=3.5$ $\mu$s for each step of gradient descent, which is the essential contribution to $\rm{TTS}$, traceable to the LLE, and represents a fundamental trade-off in power consumption and $\rm{TTS}$; see Methods.

Figure \ref{fig4}(b) explores the results of 100 full runs of the Ising machine, including time for $T_s$ evolution and readout and $N_s$ gradient-descent steps, for a single, randomly generated 3-SAT problem instance with $n=64$ variables and $m=192$ clauses. We characterize the Ising energy $H$ versus $\sum C_E$ and the histogram of $\sum C_E$ for the initial (red) and final (green) soliton spin configuration. For a 3-SAT problem, each clause has 3 variables or their negations, and thus there is a probability of 1/8 to be false for a random trial solution. Since there are 192 clauses, the number of unsatisfied clauses in the random, initial soliton spin is $\sim$24. After each complete run of the Ising machine, the final states mostly have less than 6 unsatisfied clauses with $P=5\%$ of them satisfying all clauses, i.e. $\sum C_E=0$. This demonstrates the effectiveness of the Ising machine to solve such complex, NP-complete 3-SAT problems. For random 3-SAT problems, $P$ declines with the problem size ($n$) and the clause-to-variable ratio ($m/n$) \cite{crawford1996experimental, bhattacharya2024computing}, which affects the time and energy consumed by the Ising machine to reach the solution. 

To benchmark the efficiency of 3-SAT solutions with the Kerr soliton Ising machine, we characterize $P$, $\rm{TTS}$, and the energy to solution ($\rm{ETS}$), which are standard metrics convertible between a wide range of 3-SAT solvers. We define $\rm{TTS}$ as the time to reach $\sum C_E=0$ with 99\% probability,
\begin{equation}
    \mathrm{TTS}=T_s N_s \frac{\log(1-0.99)}{\log(1-P)}.
\end{equation}
This measure of $\rm{TTS}$ considers the evolution time of solitons and excludes the calculation time of the feedback and the time to update the feedback. The other metric is energy-to-solution ($\rm{ETS}$), which is defined as the energy consumed during the $\rm{TTS}$. For a probabilistic 3-SAT solver, these metrics depend on problem size, hence the requirement for $N$, and problem complexity.
In Fig. \ref{fig4}(c), we show $P$, $\rm{TTS}$, and $\rm{ETS}$ as a function of $N=$ 32, 64, 128, 192 and 256, and $m/n=$ 2 (orange) and 3 (blue), respectively. For each $N$ and $m/n$, we generate five, random 3-SAT problems and run the problems 100 times, using the number of gradient-descent steps specified in Methods. The success probability (Fig. \ref{fig4}(c)) varies slightly, as indicated by the intervals on the data markers, for different 3-SAT instances, and it depends substantially on both $N$ and $m/n$. For $m/n=3$ (blue dots), $P$ decreases exponentially, while for $m/n=2$ (orange dots), $P$ decreases slowly for $N\leq 128$ because these problems are relatively simple to solve, and decays more rapidly for $N>128$. 
A lower success probability $P$ implies longer time and higher energy consumption to reach the solution.
The middle and lower panels of Fig. \ref{fig4}(c) present the corresponding $\rm{TTS}$ and $\rm{ETS}$, respectively. Indeed, $\rm{TTS}$ and $\rm{ETS}$ increase with $m/n$ and $N$.
We fit the data points to $e^{\sqrt{N}}$ in the dashed lines, and the agreement between the data and the fitting indicates that the Ising machine has an advantage over the general algorithm to solve the Ising model \cite{leleu2021scaling}. 

Although there is no universal algorithm for solving combinatorial optimization problems efficiently, specialized algorithms like WalkSAT can accelerate the solution of specific problems, such as 3-SAT.
We benchmark the performance of our Ising machine by comparing it with the WalkSAT algorithm, since randomized algorithms like WalkSAT with good implementations on modern computing hardware can be much faster than other heuristic methods such as simulated annealing \cite{selman1994noise}.
We implement the WalkSAT algorithm on a computer and calculate its $\rm{TTS}$ and $\rm{ETS}$ for $m/n=3$, and plot them in the black curve in the middle and lower panels of Fig. \ref{fig4}(c).
It shows that our Kerr-soliton Ising machine, which is designed to solve general Ising models, still takes less or the same time and consumes 2 orders of magnitude less optical energy than the WalkSAT algorithm. 

\section*{Conclusion}

We have demonstrated a Kerr-soliton Ising machine capable of solving combinatorial optimization problems, notably 3-SAT, with high speed and low energy consumption. Our system leverages the intrinsic bistability of Kerr microresonator solitons to encode spin states and employs a hybrid optoelectronic feedback network for fully programmable, all-to-all interactions among up to 256 spins. Each soliton bifurcates into one of two stable amplitude states under tailored feedback, providing a robust physical mechanism to emulate discrete Ising spins. By implementing gradient descent, we solve Ising-encoded SAT instances with step times as short as 3.5 µs and per-soliton power consumption of just 2.3 mW.

We benchmark the Kerr-soliton Ising machine by use of randomly generated 3-SAT problems and compare its performance against WalkSAT, a very effective heuristic solver at these problem sizes. Despite its analog nature, our system achieves comparable or better time-to-solution with two orders of magnitude lower energy use. These results underscore the potential of Kerr solitons as nonlinear, energy-efficient primitives for optical computation and position our architecture as a promising candidate for solving large-scale NP-complete problems.

Looking forward, the Kerr-soliton platform offers unique opportunities for photonic integration and scaling. Emerging foundry-compatible microresonator designs and monolithic integration of pulse generators and modulators on lithium niobate could dramatically reduce system overhead. With electronic feedback circuits operating at 20 Gbps and longer resonators or faster modulation, soliton ensembles could scale to many thousands of interacting spins. These advances would unlock ultrafast, analog optical co-processors capable of accelerating optimization and inference tasks in machine learning, cryptography, and scientific computing.

\section*{Acknowledgment}
We thank Lindell Williams and Sarang Yeola for technical review of the paper and the DARPA QuICC program management team for helpful comments throughout the experiments. This research has been funded by the DARPA QuICC program, AFOSR FA9550-20-1-0004 Project Number 19RT1019, and NIST. This work is a contribution of the U.S. government and is not subject to copyright. Trade names provide information, not an endorsement. Mention of commercial products used in the research is for informational purposes only and does not represent an endorsement.

\section*{Methods}

\textbf{Lugiato-Lefever Equation.} Here we introduce the Lugiato-Lefever Equation (LLE) with the feedback. For an ensemble of $N$ solitons, the LLE has the form of 
\begin{multline}\label{eq3}
    \frac{d a(\theta_i)}{d\tau} = -(1+i\alpha)a(\theta_i) + i\frac{d_2}{2}\frac{\partial^2}{\partial \theta^2} a(\theta_i) \\
    + i|a(\theta_i)|^2 a(\theta_i) + f(\theta_i),
\end{multline}
where $\theta_i\in [2\pi i/N, 2\pi (i+1)/N]$, $a(\theta_i)$ is the $i$-th soliton, $\alpha$ is the detuning, $d_2$ is the normalized second-order dispersion coefficient and $|f(\theta_i)|^2=|f_0(\theta_i)|^2(1+\xi_i)$ is the pump pulse where $\xi_i$ is the feedback and $f_0(\theta_i)$ is the preset pump waveform. To prevent $\xi_i$ from exceeding the limits of the AWG's input range, we truncate $\xi_i$ to the range [-1, 1] via computer control.
The waveform $f_0(\theta_i)$ is chosen such that when $\xi_i=0$, $|a_i|\simeq|a_0|$. 

\textbf{Solutions to the 3-SAT problem from an Ising problem.}  We apply the method in \cite{QUBO} to translate the 3-SAT problem with $n$ variables ($V_1$ ... $V_n$) and $m$ clauses ($C_1$ ... $C_m$) to a QUBO problem, and then convert the QUBO problem to an Ising problem with $N=m+n$ spins. After the Ising machine evolves to the final state, we obtain a spin array with $N$ spins, and the first $n$ spins are the solutions to the 3-SAT problem, i.e., $\sigma_i=1 \longleftrightarrow V_i=\rm{True}$, $\sigma_i=-1 \longleftrightarrow V_i=\rm{False}$.

\textbf{Variation of pump power.} 
The variation in pump power observed in Fig. \ref{fig3}(d) can be attributed to the EDFA's automatic gain control.
When the pulses first enter the EDFA, they are amplified maximally due to the absence of preceding pulses, resulting in a higher peak power.
As more pulses arrive, the EDFA adjusts its gain to maintain a constant average power, leading to a gradual decrease in the peak power of the pump pulses.
This, in turn, affects the soliton pulses, which initially build up and reach a maximum power before slightly decreasing with the pump pulses.
After the pump pulses stop at the 47th round trip, the soliton peaks briefly increase due to the absence of interference from incoming pump pulses, but soon decay exponentially due to the lack of pump power.
We note that this effect does not impact the Ising machine's ability to solve combinatorial optimization problems.

\textbf{Experimental operation of the Ising machine.} We operate the Ising machine with an external computer that controls the AWG and reads data from an oscilloscope. 
We form and send a pulse train to the AWG which repeatedly modulates the intensity modulator (IM2) at approximately 60 GSa/s, and read the soliton state from the oscilloscope. 
Each pulse in the pulse train has a width of 135 ps. 
To calibrate the Ising machine, we first send a train of pulses with increasing intensity and measure the corresponding soliton intensity.
By fitting the soliton intensity to the pump intensity, as shown in Fig. \ref{fig2}(b), we determine the pulse height $|f_0(\theta_i)|^2$ that corresponds to zero feedback ($\xi=0$).
Based on the height, we initiate the Ising machine with pump pulses of uniform height, update the pulse height with the feedback $\xi$, and record the resulting soliton states.
Each step takes around 0.1 s, and most of the time is taken to send the pulse to the AWG and read the soliton trace from the oscilloscope.
See Supplementary Information for more details on the benchmark problems we have solved and the feedback network.

\textbf{Number of steps for different $N$ and $m/n$.} For problems of different size and different clause-to-variable ratio, we apply different numbers of steps to reach a reliable success probability in a reasonable time.  For the problems in Fig. \ref{fig4}(c), We apply $N_s=$ 100, 100, 100, 50, 50 for $N=$ 256, 192, 128, 64, 32 and $m/n=$ 3, and $N_s=$ 100, 100, 75, 50, 50 for $N=$ 256, 192, 128, 64, 32 and $m/n=$ 2, respectively.

\textbf{Numbers of variables and clauses for different spin numbers} For the spin number $N$ which is a power of 2, $m/n$ cannot exactly equal to 2. For a given $N$, we choose the $m$ and $n$ such that $m/n$ is the closest to 2 while $m/n\geq 2$. With $m/n=2$, $m=22$ and $n=10$ for $N=32$, $m=43$ and $n=21$ for $N=64$, $m=86$ and $n=42$ for $N=128$, $m=128$ and $n=64$ for $N=192$, $m=171$ and $n=85$ for $N=256$.

\textbf{Threshold of the solitons.} For continuous-wave pumping, the expression for the threshold of the solitons in a Kerr resonator is
\begin{equation}
    P_{th} = \frac{\pi n_g^2 LA_{\mathrm{eff}}}{4 n_2\lambda Q_i^2} \frac{(1+K)^3}{K}
\end{equation}
where $Q_i$ is the intrinsic quality factor, $Q_e$ is the external quality factor, $K=Q_i/Q_e$ is the coupling coefficient, $n_g$ is the group index, $n_2$ is the nonlinear index, $L$ is the circumference of the resonator, $A_\mathrm{eff}$ is the fiber effective area, and $\lambda$ is the wavelength. 
In our experiment, $Q_i=1.34\times 10^9$, $Q_e=4.61\times 10^8$, $n_g=1.44$, $A_{\rm{eff}}=8.5\times 10^{-11}$ m$^2$, $L=29$ m, $n_2=3.11\times 10^{-20}$ m$^2$/W and $\lambda=1556$ nm.
The circumference $L$ of the resonator includes the length of the SMF (25 m) and the couplers (around 2 m each).
For the fiber resonator that we use, the predicted threshold is $P_{th}=0.9$ W for continuous wave pumping. 
Since we use a pulse pump and the width of a single pulse is around 1/2048 of a round trip, the threshold to generate a single soliton is around 0.45 mW, which is comparable to 2.3 mW measured in the experiment considering the loss after the EDFA. Use of a microcomb to generate the pulse pump would be ideal, enabling $P_{th}\sim 0.05$ mW. As we increase the number of pump pulses to activate a larger soliton ensemble, $P_{th}$ is constant. 

\textbf{$\rm{TTS}$ and $\rm{ETS}$ with respect to $K$.} Here we explore the relation of $\rm{TTS}$ and $\rm{ETS}$ with the coupling coefficient $K$. Since the normalized LLE \eqref{eq3} does not directly include $K$, $\rm{TTS}$ is fundamentally determined by the time unit $2/\kappa$. Thus we have
\begin{equation}
    \mathrm{TTS}\propto \frac{2}{\kappa} \propto \frac{\lambda Q_i}{1+K}.
\end{equation}
Since the $\rm{ETS}$ is the energy consumed during $\rm{TTS}$, we have
\begin{equation}\label{ETS}
    \mathrm{ETS} \propto \mathrm{TTS}\times P_{th}\propto \frac{1}{Q_i}\frac{(1+K)^2}{K}.
\end{equation}
In our system, $Q_i$ comes from the loss of splicing between fibers and the excess loss of the couplers, while $Q_c$ and $K$ can be tuned by changing the coupling ratio of the couplers. From Eq. \ref{ETS} the $\rm{ETS}$ reaches the minimum with $K=1$.

\section*{Data Availability}

Source Data will be provided with this paper. Further data are available from the corresponding author on reasonable request.

\printbibliography

\end{document}